\documentstyle[11pt]{article}

\input{hepsfig.sty}

\begin{document}
\baselineskip=6mm

                                       %  to Chinese Physics Letters
                                       %  1998.03.24

\begin{center}
{\Large\bf Fireball/Blastwave Model and Soft $\gamma$-ray Repeaters }
\footnote{ Supported by the National Natural Science Foundation 
of China under grants 19573006 and 19773007, 
and the Foundation 
of the Ministry of Education of China.}

\vspace{5.0mm}
HUANG Yong-feng, \hspace{1mm} Dai Zi-gao, \hspace{1mm}  LU Tan

{\sl Department of Astronomy, Nanjing University, Nanjing 210093}

\end{center}

\vspace{5mm}
\noindent
{\Large\it Already published in: \\ {\Large\it\bf Chinese Physics Letters, 1998, 15, 775}} 

\vspace{5mm}
\noindent
{\footnotesize\sl
Soft $\gamma$-ray repeaters are at determined distances
and their positions are known accurately. If observed, afterglows from 
their soft $\gamma$-ray bursts will provide important clues to the  
study of the so called ``classical $\gamma$-ray 
bursts''. On applying the popular fireball/blastwave model of 
classical $\gamma$-ray bursts to soft $\gamma$-ray repeaters, it
is found that their X-ray and optical afterglows are detectable.
Monitoring of the three repeaters is solicited.  }

\vspace{5mm}

\noindent
{\it PACS: \hspace{1mm} 98.70.Rz, 98.70.Qy, 97.60.Jd } 
% Gamma-ray sources 
% X-ray sources 
% Neutron stars

\vspace{5mm}

Since their discovery nearly thirty years ago, $\gamma$-ray bursts (GRBs) 
have made one of the biggest mysteries in astrophysics, primarily 
because they have remained invisible at wavelengths other than
$\gamma$-rays so that the distances are unknown.$^{1}$ 
The recent detection of X-ray, optical and even radio afterglows from some
GRBs located by the Italian-Dutch BeppoSAX satellite has opened  up a new 
era.$^{2,3}$ The possible host galaxy of GRB 970228 and the determined 
redshift $0.835 < z < 2.1$ for GRB 970508 strongly indicate a cosmological 
origin. Fireball model becomes the most popular and successful 
model for GRBs. After the main GRB, the blastwave generated between the GRB 
ejecta and the interstellar medium (ISM) provides a natural explanation for 
the power-law decay of the observed low energy afterglows.$^{2-4}$ 
We call such a
GRB scenario as a fireball/blastwave model. However, so few GRBs have
been located rapidly and accurately enough for us to search for their
afterglows, that the cosmological origin of GRBs and the correctness
of the fireball/blastwave model still need more tests. GRBs 
occurring at a definite distance and in a fixed direction would be ideal
for checking the model. Here we suggest that soft $\gamma$-ray
repeaters (SGRs), whose nature is much clearer, might be good candidates.

As a subtle class of GRBs, SGRs are characterized 
mainly by their soft spectra and unpredictable recurrences.$^{5}$ 
There are only three known SGRs: 0526$-$66, 1806$-$20 and 1900$+$14, all
of them have been tentatively associated with supernova 
remnants (SNRs), indicating a neutron star origin. Recently,
Hurley {\em et al.} (1997) seem to
have observed a new (4th) soft $\gamma$-ray
repeater.$^{7}$ A typical SGR burst lasts 
several hundred milliseconds, emitting 
about $10^{39}$ $-$ $10^{41}$ erg in soft $\gamma$-rays. Due to
the huge energies, the limited volume, and the small timescale, a 
fireball seems inevitable before soft $\gamma$-rays are emitted, just
as for a cosmological GRB. This has led to our suggestion that we
could check the fireball model by monitoring the SGR sources.

In fact, let us consider a fireball with a total radiation energy
$E_{41} \times 10^{41}$ erg and a radius $r_6 \times 10^6$ cm 
in thermal equilibrium.$^{6}$ The temperature will be 
$T = 113 E_{41}^{1/4} r_6^{-3/4}$ keV. The optical depth due to 
Compton scattering of photons from e$^{\pm}$ pairs is 
$\tau = 3.0 \times 10^{11} E_{41}^{3/8} r_6^{-1/8} 
\exp (-4.5 r_6^{3/4} / E_{41}^{1/4})$, obviously optically thick for
a typical SGR burst. The same conclusion would be drawn even if the
energy was supposed to be released steadily with a luminosity 
$L > 10^{40}$ erg$\cdot$s$^{-1}$.$^{6}$
Below we will briefly describe the fireball/blastwave model and apply
it to SGR bursts to predict their afterglows in X-ray and optical bands.

A fireball with total initial energy $E_0$ and initial bulk Lorentz 
factor $\eta \equiv E_0/M_0 c^2$, where $M_0$ is the initial baryon mass
and $c$ the velocity of light, is expected to radiate half of its energy 
in $\gamma$-rays during the GRB phase, either due 
to an internal-shock or an external-shock mechanism.
Subsequently the fireball will continue to expand as 
a thin shell into the ISM, generating an ultrarelativistic shock, which
has already been studied analytically. 
A simple approximate solution for the shell radius, $R(t)$, and the 
Lorentz factor of the shocked ISM, $\gamma (t)$, is derived as:
\begin{equation}
R(t) \approx 8.93 \times 10^{15} E_{51}^{1/4} n_1^{-1/4} t^{1/4} {\rm cm} 
 = 2.82 \times 10^{13} E_{41}^{1/4} n_1^{-1/4} t^{1/4} {\rm cm},
\end{equation}
\begin{equation}
\gamma (t) \approx 193 E_{51}^{1/8} n_1^{-1/8} t^{-3/8} 
 = 10.9 E_{41}^{1/8} n_1^{-1/8} t^{-3/8},
\end{equation}
where $E_0 = 10^{51} E_{51}$ erg $= 10^{41} E_{41}$ erg, $n = n_1$ 
cm$^{-3}$ is the number density of ISM and $t$ is observer's time in units 
of 1 s.These equations are good
approaches only when $\gamma \gg 1$ and $t \gg t_{\rm G}$ (duration of the
main GRB). For a more accurate solution 
suitable even when $t \approx t_{\rm G}$
and/or $\gamma \approx 1$, please see Huang {\em et al.}'s numerical 
evaluation.$^{4}$

Electrons in the shocked ISM are highly relativistic. Inverse Compton 
cooling may not contribute to emission in X-ray and optical bands 
that we are interested in. We will consider only synchrotron
radiation below. In the comoving frame the electron number 
density ($n_{\rm e}'$) distribution in the shocked ISM is assumed
to be a power-law function of electron Lorentz 
factor $\gamma_{\rm e}$, as expected for shock
acceleration, $dn_{\rm e}' / d \gamma_{\rm e} \propto \gamma_{\rm e}^{-p}$, 
($\gamma_{\rm min} \leq \gamma_{\rm e} \leq \gamma_{\rm max}$), 
where $\gamma_{\rm min}$ 
and $\gamma_{\rm max}$ are the minimum and maximum Lorentz factors, and $p$
is the index varying between 2 and 3. We suppose that the magnetic field
energy density is a fraction $\xi_{\rm B}^2$ of the
energy density $e'$, $B'^{2}/8 \pi = \xi_{\rm B}^{2} e'$,
where $B'$ is the magnetic 
field in the comoving frame, and the electron carries a
fraction $\xi_{\rm e}$ of the energy, so that 
$\gamma_{\rm min} = \xi_{\rm e} (m_{\rm p}/m_{\rm e}) \gamma (p-2)/(p-1)$, 
where $m_{\rm p}$ and $m_{\rm e}$ are proton and electron masses 
respectively, and $\gamma_{\rm max} = 10^8 B'^{-1/2}$. 
Synchrotron radiation is then translated from
the comoving frame into the observer's frame. The derived flux densities
at frequency $\nu$ 
decay as a power-law, $S_{\nu} \propto t^{- \alpha}$, where 
$\alpha = 3(p-1)/4$, in good agreement with recent observations.

Above is only a rough depict. We have carried out detailed numerical
evaluation to investigate the 
afterglows from SGR bursts, following Huang {\em et al.}'s simple model.$^{4}$
We chose $E_0$ between $10^{40}$ and $10^{42}$ erg, 
and $n=1$ or 10 cm$^{-3}$.  In each case we
set $p=2.5$, $\xi_{\rm e} = 0.5$, $\xi_{\rm B} = 0.1$ and $d = 10$ kpc. 
Since $M_0$ is 
a parameter having little influence on the afterglows,
we chose $M_0$ so that $\eta \approx 280$ in all cases. X-ray flux 
$F_{\rm X}$ is integrated from 0.1 to 10 keV, and optical flux densities 
for R band $S_{\rm R}$ are calculated. The evolution of $F_{\rm X}$ and 
$S_{\rm R}$ are illustrated in Figs. 1 and 2 respectively. We see that for 
a strong burst ($E_0 > 10^{41}$ erg), $F_{\rm X}$ can in general keep to be
above $10^{-12}$ erg$\cdot$cm$^{-2} \cdot$s$^{-1}$ for 40 to
more than 200 s
and $S_{\rm R}$ can keep to be 
above $10^{-29}$ erg$\cdot$cm$^{-2} \cdot$s$^{-1}$ Hz$^{-1}$ 
(corresponding to magnitude $R \approx 24.0$ mag) for 200 to 
more than $10^3$ s. But if we take 
$E_0 = 10^{40}$ erg, then $F_{\rm X}$ can hardly be greater than 
$2 \times 10^{-12}$ erg$\cdot$cm$^{-2} \cdot$s$^{-1}$.

Now we discuss the observability. 
The three known SGRs have been extensively looked after in X-ray, optical
and radio bands. A pointlike X-ray source has been identified 
in each case, but only SGR 1806$-$20 has a
detectable optical counterpart.  SGR 0526$-$66 is associated with 
the Large Magnellanic Cloud SNR N49, about 55 kpc from us.
A permanent X-ray hot spot is found with an unabsorbed flux of 
about $2.0 \times 10^{-12}$ erg$\cdot$cm$^{-2} \cdot$s$^{-1}$ 
(0.1 $-$ 2.4 keV).$^{8}$ No optical counterpart brighter 
than magnitude $m_{\rm v} = 21$ mag has been identified.
SGR 1806$-$20 is associated with the Galactic SNR G10.0$-$0.3, about 10
to 15 kpc from the Earth. A steady pointlike X-ray source with an 
unabsorbed flux of 
about $10 \times 10^{-12}$ erg$\cdot$cm$^{-2} \cdot$s$^{-1}$
has been observed.$^{9}$ Optical observations have revealed a
luminous companion (spectral type O9 $-$ B2) to 
this SGR, but heavily reddened due
to serious interstellar extinction ($A_{\rm v} \approx 30$ mag). 
The least active source SGR 1900$+$14 is associated with the Galactic 
SNR G42.8$+$0.6, about 7 to 14 kpc from us. A quiescent, steady, 
point X-ray source is present at its position,$^{10}$ with
an unabsorbed flux 
of $3.0 \times 10^{-12}$ erg$\cdot$cm$^{-2} \cdot$s$^{-1}$. 
No optical source is detected down to limiting magnitude of 
$m_{\rm v} \approx 24.5$ mag.

In order to be detectable, the X-ray afterglow flux from an SGR burst 
should at least be comparable to that of the quiescent X-ray source.
Taking $10^{-12}$ erg$\cdot$cm$^{-2} \cdot$s$^{-1}$ as a threshold, then the
predicted afterglows will generally be above the value for about
40 $-$ 200 s for intense events (Fig. 1). 
Since the peak flux can be as high
as $10^{-8}$ $-$ $10^{-7}$ erg$\cdot$cm$^{-2} \cdot$s$^{-1}$, such an afterglow
should be observable by those satellites now in operation, such as 
ROSAT, ASCA and Rossi. If detected afterglows from SGR bursts would be
ideal to test the fireball/blastwave model, so we suggest that SGRs 
should be monitored during their active periods. Cases are similar for optical 
afterglows. If we took $S_{\rm R} = 100$ $\mu$Jy ($R \approx 19$ mag) as
the threshold, afterglow would last less than 100 s, but if we took
$S_{\rm R} = 1$ $\mu$Jy ($R \approx 24$ mag), then we would have more 
than $10^3$ s (Fig. 2).

We notice that some researchers do have monitored the SGRs in optical
and radio wave bands. In 1995, Vasisht {\em et al.} reported a negative 
detection by the Very Large Array of any radio variability from SGR 1806$-$20 
above the 25\% level on postburst timescales ranging from 
2 d to 3 month.  Radio afterglows are beyond our discussion here 
because strong self-absorption is involved.
In 1984, Pedersen {\em et al.} reported three possible optical flashes 
from SGR 0526$-$66, but none of their light curves shows any sign of 
afterglows. We think,they were either due to the limited aperture (50 cm)
of their telescope or maybe simply spurious. The latter
seems more possible since no soft $\gamma$-ray bursts were observed 
simultaneously with them.

Of special interest is the most prolific source SGR 1806$-$20. In 
1993 October 9.952414 UT a soft $\gamma$-ray burst was
recorded by the Compton Gamma-Ray Observatory. ASCA 
satellite happened to be observing the SGR at that moment and 
a coincident X-ray burst was recorded.$^{9}$
Sonobe {\em et al.} particularly pointed out that there were
no obvious mean intensity changes in X-rays prior to the burst nor 
following the burst, not only on a timescale of 1 d, but also on 
timescales of minutes.$^{11}$ This is not inconsistent with our model 
since it was a relatively weak burst, with $E_0$ about $10^{39}$ 
erg. Afterglows from this burst are not expected to be detectable.

We have also calculated the afterglows from such a unique burst as 
GRB 790305 from SGR 0526$-$66,$^{12}$ 
taking $E_0$ to be $1 \times 10^{45}$ erg and $d = 55$ kpc.
We find that the X-ray afterglows should be 
detectable ($> 10^{-12}$ erg$\cdot$cm$^{-2} \cdot$s$^{-1}$) for
several hours, and $S_{\rm R}$ 
will be above 100 $\mu$Jy ($R \approx 19$ mag) for about one hour.
Had the source been monitored on 1979 March 5, chances were good that
afterglows should have been observed.

We end this letter by a brief comment on the importance of our 
suggestion. GRBs occurring at three different distance 
scales have been observed or suggested: classical GRBs at cosmological 
distances, classical GRBs in the Galactic Halo, and SGRs at about
10 kpc distances. Classical GRBs are intriguing puzzles because, if
occurred at cosmological distances, they would present stringent 
requirements on the energies and the initial baryon masses. 
The cosmological origin of classical GRBs and the fireball/blastwave
model are two propositions. Although they are consistent with each other
in that the observed power-law decays of afterglows from GRBs can be
naturally explained, both of them are in urgent need of more observational
tests, especially independent ones. The possible host galaxy of 
GRB 970228 and the red shift of GRB 970508 are two strong proofs,
but far from enough. Here we have suggested that the three known 
SGRs, especially the most prolific one, SGR 1806$-$20, are good 
candidates to be used to test the fireball/blastwave model independently.
Our arguments are obvious: the distances are much certain, their
accurate positions are available, they burst out repeatedly, their
origins are relatively clear so that we feel more confident about them. 
Although such monitoring observations are imaginably difficult, the
results will be valuable, not only in that the afterglows might be 
acquired, but also that the simultaneous bursting behaviors in 
X-ray and optical wavelengths other than soft $\gamma$-rays are
important to our understanding of these SGRs themselves.

\vspace{10mm}

{\noindent\bf
REFERENCES}

\vspace{5mm}

\begin{description}
\baselineskip=2mm
\item {[1] G. J. Fishman and C. A. Meegan, Ann. Rev. Astron. Astrophys. 
       33 (1995) 415.}
\item {[2] E. Costa {\em et al.}, Nature 387 (1997) 783.} 
\item {[3] S. G. Djorgovski {\em et al.}, Nature 387 (1997) 876.}
\item {[4] Y. F. Huang, Z. G. Dai, D. M. Wei and T. Lu, 
       Mon. Not. R. Astron. Soc. (1998) in press.}
\item {[5] J. P. Norris, P. Hertz, K. S. Wood and C. Kouveliotou, 
       Astrophys. J. 366 (1991) 240.} 
\item {[6] T. Piran and A. Shemi, Astrophys. J. 403 (1993) L67.}
\item {[7] K. Hurley, C. Kouveliotou, T. Cline and E. Mazets,
       IAU Circular, No.6743 (1997).}
\item {[8] D. Marsden, R. E. Rothschild, R. E. Lingenfelter and 
       R. C. Puetter, Astrophys. J. 470 (1996) 513.}
\item {[9] T. Murakami {\em et al.}, Nature 368 (1994) 127.}
\item {[10] K. Hurley {\em et al.}, Astrophys. J. 463 (1996) L13.}
\item {[11] T. Sonobe, T. Murakami, S. R. Kulkarni, T. Aoki and 
       A. Yoshida, Astrophys. J. 436 (1994) L23.}
\item {[12] W. D. Evans {\em et al.}, Astrophys. J. 237 (1980) L7.}

\end{description}

%\vspace{25mm}

\begin{figure}[htb]
  \begin{center}
  \leavevmode
  \centerline{ 
  \epsfig{figure=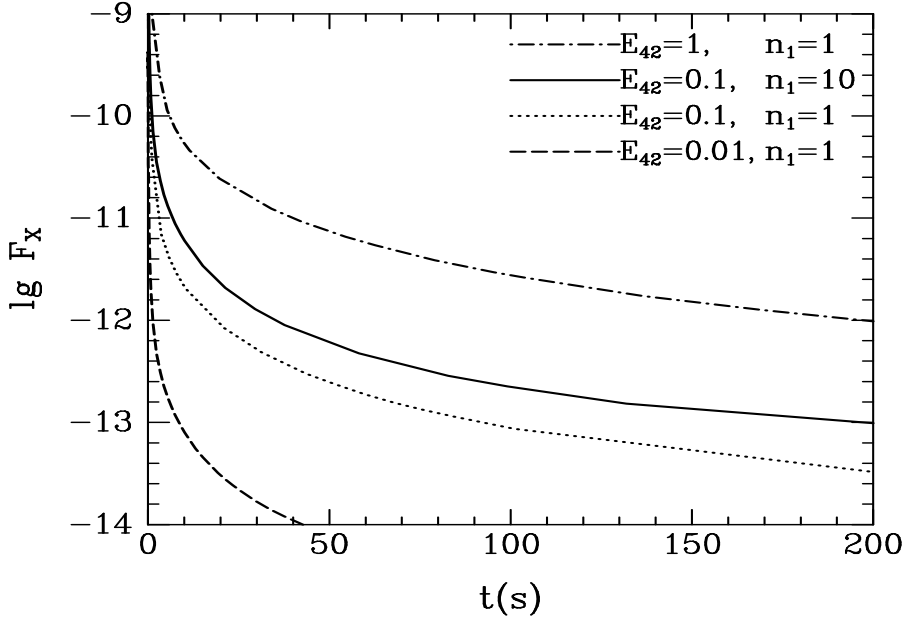,width=3.0in,height=3.0in,angle=270,
  bbllx=150pt, bblly=200pt, bburx=525pt, bbury=550pt}
  }
\caption {Predicted X-ray afterglows from SGR bursts.
Flux $F_{\rm X}$ is integrated from 0.1 to 10 keV and is in unit of
erg$\cdot$cm$^{-2} \cdot$s$^{-1}$. Time is
measured from the end of the main $\gamma$-ray burst. }
  \end{center}
  \end{figure}

\begin{figure}[htb]
  \begin{center}
  \leavevmode
  \centerline{ 
  \epsfig{figure=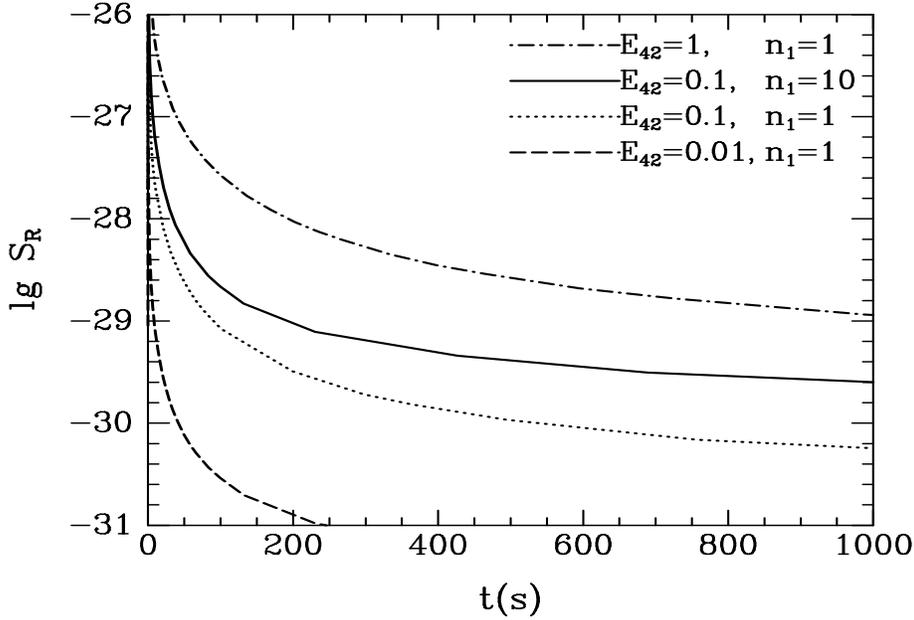,width=3.0in,height=3.0in,angle=270,
  bbllx=150pt, bblly=200pt, bburx=525pt, bbury=550pt}
  }
\caption { Predicted optical afterglows from SGR bursts. 
The R band flux density $S_{\rm R}$ is in unit of
erg$\cdot$cm$^{-2} \cdot$s$^{-1} \cdot$Hz$^{-1}$.}
  \end{center}
  \end{figure}

\end{document}